\documentclass[12pt]{iopart}

\usepackage{graphicx}
\usepackage{braket}
\usepackage{bbold}
\usepackage[english]{babel}
\usepackage{cite}

\usepackage{iopams}  
\begin{document}   

\title{Two-photon interference at a telecom wavelength for quantum networking}

\author{Mathis Cohen, Laurent Labonté, Romain Dalidet, Sébastien Tanzilli, Anthony Martin}

\address{Université Côte d’Azur, CNRS, Institut de physique de Nice, France}
\ead{mathis.cohen@univ-cotedazur.fr}
\vspace{10pt}

\begin{abstract}
 The interference between two independent photons stands as a crucial aspect of numerous quantum information protocols and technologies. In this work, we leverage fiber-coupled devices, which encompass fibered photon pair-sources and off-the-shelf optics, to demonstrate Hong-Ou-Mandel interference. We employ two distinct single photon sources, namely an heralded single-photon source and a weak coherent laser source, both operating asynchronously in continuous-wave regime. We record two-photon coincidences, showing a state-of-art visibility of 91.9(5)\%. This work, compliant with telecom technology provides realistic backbones for establishing long-range communication based on quantum teleportation in hybrid quantum networks.   
\end{abstract}

\section{Introduction}
Quantum networks represent the backbone of the future quantum internet, seamlessly connecting remote quantum devices, such as quantum sensors, processors, or users, via photonic quantum resources producing single and/or entangled bits on various observables~\cite{labonte2024integrated}. Such networks should significantly advance various application scenarios such as distributed quantum computing~\cite{munro_designing_2022}, device-independent quantum key distribution~\cite{PhysRevLett.98.230501, Pironio_2009}, and distributed quantum sensing~\cite{chen_quantum_2022}.

\begin{figure}[!ht]
    \centering
    \includegraphics[width=0.9\linewidth]{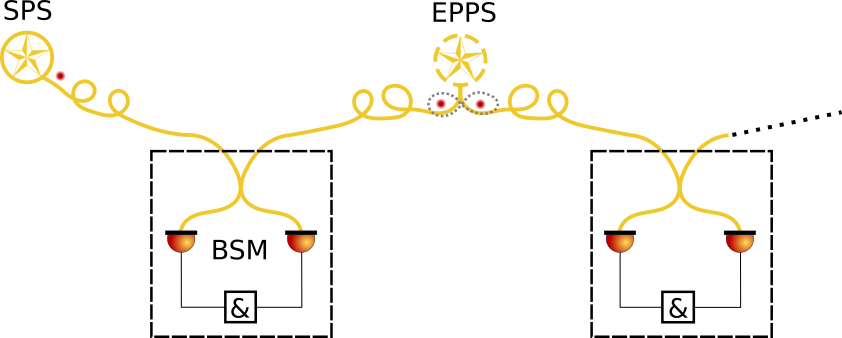}
    \caption{Simple schematic of a quantum network operation for long distance propagation, using a single-photon source (SPS), an entangled photon-pair source (EPPS), and a Bell state measurement (BSM) apparatus.}
    \label{Principe}
\end{figure}

Quantum teleportation between photonic qubits stands as the centerpiece of this framework. It plays a critical role in enabling the formation of entanglement between two remote sites without requiring single photons traveling the overall distance, as illustrated in Fig.\ref{Principe}~\cite{PhysRevLett.126.130502, pfaff_unconditional_2014}. By harnessing Bell-state measurements to combine different qubits (originating from various types of sources), as in quantum relays and repeaters, qubits that never physically interact can become entangled~\cite{valivarthi_teleportation_2020, takesue_quantum_2015, xia_long_2017, samara_entanglement_2020}. Through this chain of operations, entanglement swapping permits the creation of end-to-end quantum links between users arbitrarily separated in space, regardless of the nature of the photon sources. The condition to successful implementation of these operations relies on the ability to achieve high-visibility two-photon interference, measured using the so-called Hong-Ou-Mandel (HOM) experiment. In such a realization, the obtained visibility attests the indistinguishability of the interfering photons in all degrees of freedom, including polarization, spectral, time, and spatial modes.

Telecom photons stand as ideal carriers for transporting qubits over relatively long distances using standard optical fibers. Also, filtering and routing these photons can be handled advantageously using off-the-shelf fiber components~\cite{tanzilli_photonic_2005}. These carriers can have different origins, including sources of entangled photons, single photons (SPS), as well as weak coherent laser pulses (WCS). The origin of the photons does not matter to perform a HOM type experiment, as only the photonic properties of the produced single photons need to be considered in order to optimize the interference pattern. 

To date, the HOM effect has been extensively studied using the same generation process, including second-order $\chi^{(2)}$ and third-order $\chi^{(3)}$ nonlinear crystals~\cite{mcmillan_two-photon_2013, rarity_high-visibility_1993, martin_quantum_2012}, quantum dots~\cite{PhysRevLett.104.137401, patel_two-photon_2010, santori_indistinguishable_2002}, NV centers in diamond~\cite{PhysRevLett.85.290}, single atoms~\cite{beugnon_quantum_2006}, atomic ensembles~\cite{felinto_conditional_2006}, trapped ions~\cite{maunz_quantum_2007}, and weak coherent pulses~\cite{Kim:20}. Within the contemporary context of hybrid quantum networks, HOM experiments built with different types of sources, having each complementary and distinct roles, have to be considered~\cite{mcmillan_two-photon_2013, huwer_quantum-dot_2017, shields_entangled-led-driven_2016}. A pioneering experiment based on the interference between weak coherent pulses and single photon states, both synchronized by a master laser, has demonstrated a HOM visibility of 60\%~\cite{rarity_non-classical_2005}. In this work, The remaining challenge is to achieve optimal mode overlap across all degrees of freedom, particularly the spectral/temporal one, by ensuring that the coherence time of the two interfering photons is comparable to, or longer than, the detection timing jitter. Furthermore, this approach does not address the issue of scalability, as a common master clock is needed to ensure synchronization of the temporal modes and to eliminate residual phase drifts.

Here, we carry out two-photon interference (TPI) between weak coherent and single photon states originating from a WCS and an heralded single-photon source (HSPS), respectively. Our approach is demonstrated by operating both sources in continuous-wave regime allowing for asynchronous operation, i.e. free from sharing any common master clock. Synchronization is achieved \textit{a posteriori} by detecting the heralding photon. Through precise spectral and temporal shaping of the wave packets, we achieve a raw visibility exceeding 90\% in coincidence trace. Our architecture employs one member of the photon pair to gate an intensity modulator (IM), which then drives the WCS. This strategy significantly reduces photonic noise from the continuous photon flux of the WCS, enabling gated operation. Additionally, this solution eliminates the need for triggered single-photon detectors, allowing the use of superconducting nanowire single-photon detectors (SNSPDs), which offer high detection efficiency, low noise, and a high dynamic range. This advancement greatly enhances the development of network-type protocols, such as quantum teleportation, leveraging both similar~\cite{mcmillan_two-photon_2013,valivarthi_teleportation_2020, takesue_quantum_2015, xia_long_2017, samara_entanglement_2020} and different ~\cite{huwer_quantum-dot_2017, shields_entangled-led-driven_2016} types of photon sources, thereby paving the way for more scalable and versatile quantum communication networks.

\section{Analytical model and numerical simulation}
\label{section:Model}

\begin{figure}[!ht]
    \centering
    \includegraphics[width=0.6\linewidth]{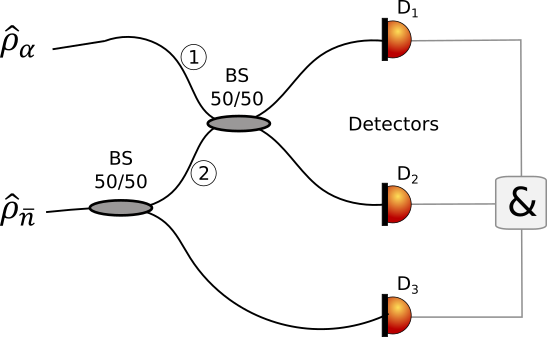}
    \caption{Model-scheme experiment. The coherent state interferes with one member of a photon pair. The separation of paired photons and the combination of the interfering are ensured by beam splitter (BS) operation. Then, detection is performed (D$_{1-3}$) and heralded two-fold coincidences are recorded (\&). 1 and 2 represent the impinging modes at the BS responsible for the HOM interference.}
    \label{ssimu}
\end{figure}

The general scheme modeling our experiment is represented in Fig.~\ref{ssimu}. We aim at determining the optimal operation regime for the sources in order to maximize the visibility of the TPI. The sources are a WCS and an HSPS, whose emission characteristics can be modeled by two parameters, the coherent state amplitude $\alpha$ and the mean photon pair number $\bar{n}$, respectively.

The overall density matrix, taking into account the produced states, can be expressed as:
\begin{equation}
     \rho_{tot}(\alpha,\bar{n}) = \rho_{\alpha}(\alpha) \otimes \rho_{\bar{n}}(\bar{n}),
\end{equation}
where $\rho_{\alpha}(\alpha)$ and $\rho_{\bar{n}}(\bar{n})$ refer to density matrices for the WCS and HSPS, respectively.

\begin{figure}[!ht]
    \centering
    \includegraphics[width=1.1\linewidth]{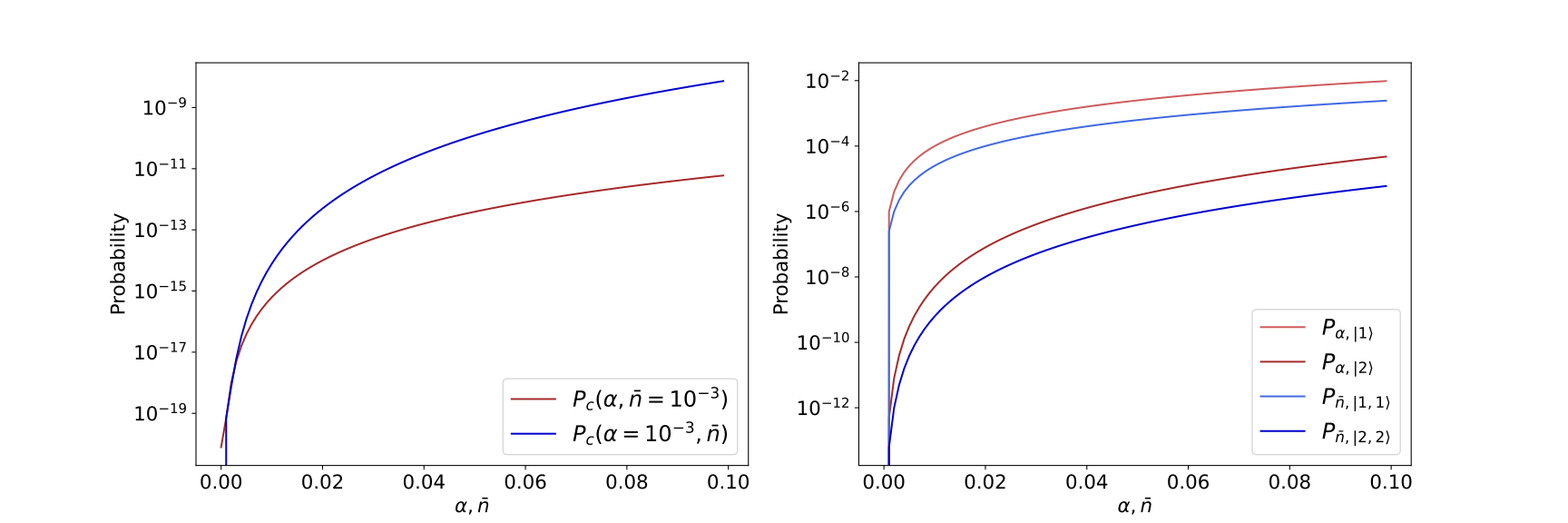}
    \caption{\small{\textit{Left}: 3-fold coincidence probability. The red curve refers to the probability as a function of $\alpha$ with $\bar{n}$ constant ($\bar{n} = 10^{-3}$). The blue curve represents the reciprocal case, with $\alpha$ constant ($\alpha = 10^{-3}$). \textit{Right}: Emission probability of the two sources depending on emission parameters $\alpha$ and $\bar{n}$. The probability associated with the three first Fock states are represented for each source.}}
    \label{prob}
\end{figure}

Based on the detection scheme presented in Fig.~\ref{ssimu}, Fig.~\ref{prob} shows the evolution of the 3-fold coincidence and emission probabilities of the two sources as a function of $\alpha$ and $\bar{n}$. Simultaneous multi-photon emission naturally increases with these parameters. This leads to a reduction in the HOM visibility by adding accidental coincidences.

To simulate interferences, we implement the BS operator written as:

\begin{equation}
     \hat{S} = e^{(i\theta (\hat{a_1}^\dag\hat{a_2} + \hat{a_1}\hat{a_2}^\dag))},
\end{equation}
 where $\theta = \pi/4$ for a 50/50 BS and $\hat{a_1}^\dag,\hat{a_2}^\dag$ are the creation operators in modes 1 and 2 of the second BS, respectively (see Fig.~\ref{ssimu}). The density matrix at the output of this BS can be expressed as:

\begin{equation}
     \rho_{BS,out}(\alpha,\bar{n}) = \hat{S}^\dag\rho_{tot}(\alpha,\bar{n})\hat{S}.
\end{equation}
The positive operator-valued measurement (POVM) method is employed to model the detection process, and is represented as $\hat{D} = (\mathbb{1} - \ket{0}\bra{0})^{\otimes 3}$. This is used to compute the number of 3-fold coincidences $N_c$, taking into account that the detectors do not resolve photon numbers. The states are combined at the second BS with the upper member of the photon pair (see Fig.~\ref{ssimu}). Therefore, the 3-fold coincidence operator is applied as follows:
\begin{equation}
     N_c^{indis} = Tr(\hat{D}^\dag\rho_{tot,out}(\alpha,\bar{n})\hat{D}).
\end{equation}
The final step introduces distinguishability between the states to calculate the visibility of the HOM interference. To simulate maximum distinguishability, the input state is decomposed such that each photon arrives at the BS separately, thereby preventing two-photon interference. The output state is subsequently reconstructed at the BS exit, and the probability of three-fold coincidence, denoted  $N_c^{dis}$, is computed. The resulting HOM visibility is then obtained by comparing the perfect case and the latter. It comes :
\begin{equation}\label{vis}
     V_{HOM} = 1 - \frac{N_c^{indis}}{N_c^{dis}}.
\end{equation}

\begin{figure}[!ht]
    \centering
    \includegraphics[width=0.9\linewidth]{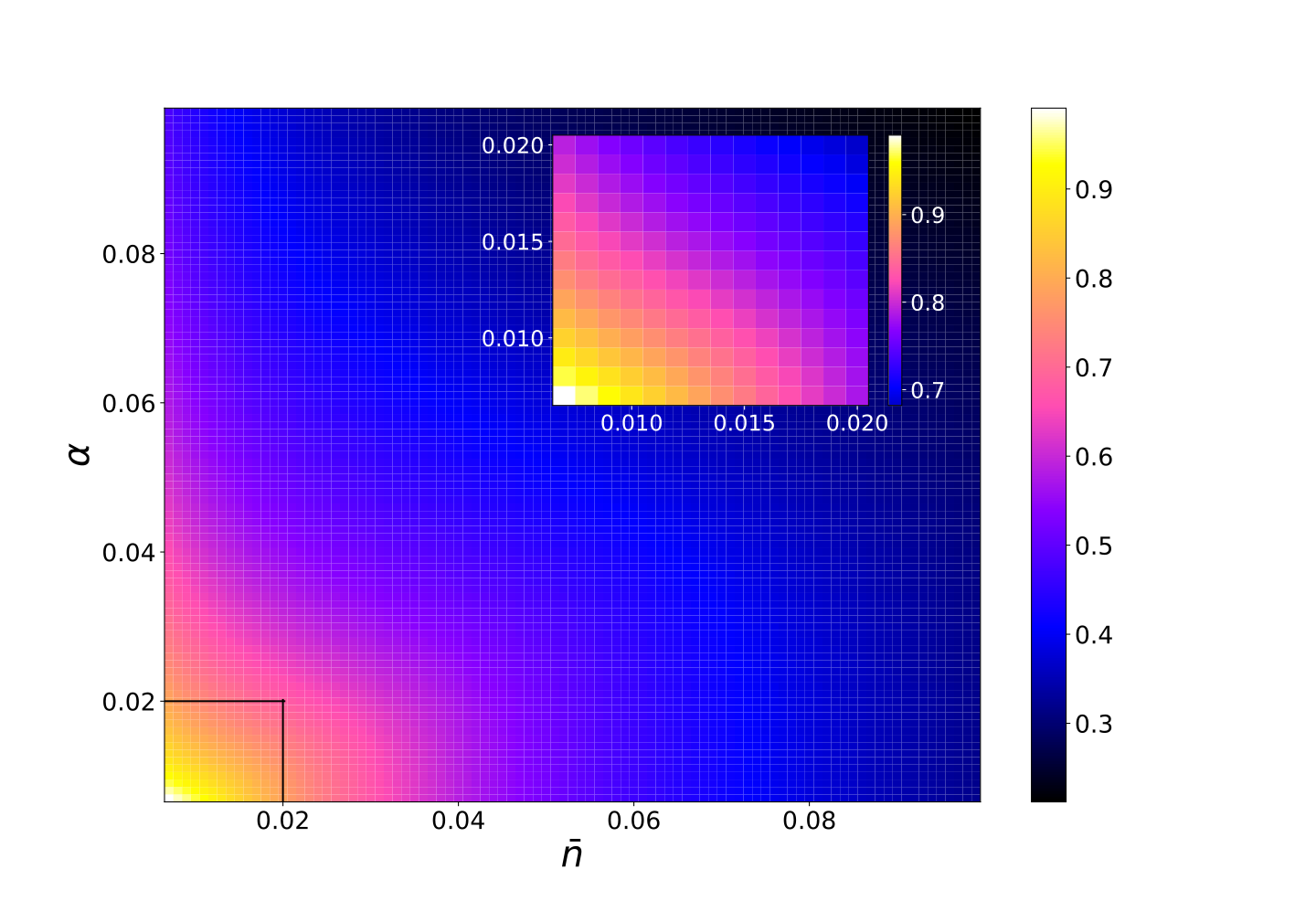}
    \caption{Simulation of the visibility of a HOM experiment between independent photons as a function of their emission parameters, $\alpha$ and $\bar{n}$. \textit{Inset}: zoom on the zone for which $\bar{n}$ and $\alpha$ are below 0.02, corresponding to a visibility above 70\%.}
    \label{simu}
\end{figure}

Fig.~\ref{simu} depicts the result obtained from the simulation, which depends on the emission parameters of the two sources. As expected, visibility decreases rapidly as the values of $\alpha$ and $\bar{n}$ increase. This simulation provides two major insights. First, the influence of the two parameters is quite similar, and the visibility mainly depends on the statistical distributions of the single photons. Secondly, it allows quantifying the values of $\bar{n}$ and $\alpha$ that must be below 0.02 to ensure visibility values above 70\%. Therefore, a trade-off is necessary, i.e. the emission regimes should be adjusted around this threshold while keeping coincidence rates at acceptable levels to effectively execute any HOM interference-based protocol within reasonable integration times.

\section{Experiment}

\begin{figure}[!ht]
    \begin{center}
        \includegraphics[width=1\linewidth]{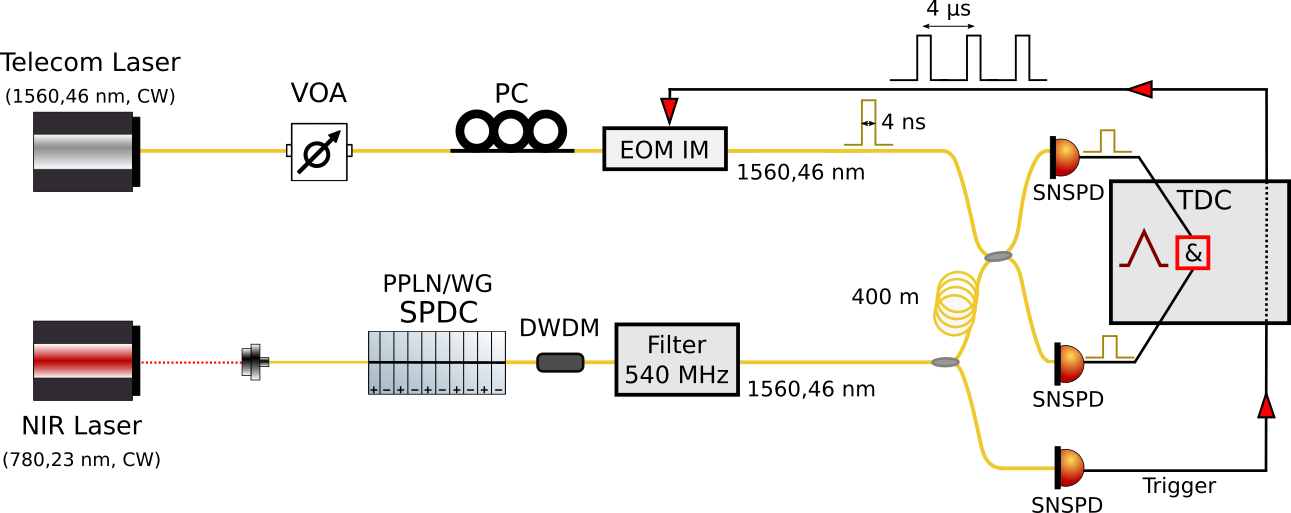}
        \caption{\small{Experimental setup of Hong-Ou-Mandel interference. The coherent state is attenuated through a variable optical attenuator (VOA) and passed through a polarization controller (PC). An intensity modulator (IM) generates WCP in the coherent state path. The trigger is sent to the IM with the detection of one member of the photon pair using superconducting nanowire single photon detectors (SNSPD), at a rate of about $250$ kHz, passing through the TDC to condition the signal (in amplitude and shape). The IM generate square optical pulses, with a width of $4$~ns. Photon detection is accomplished using SNSPDs, synchronized with respect to the trigger one, and the temporal coincidence window corresponds to the convolution of temporal signals received by the detectors, here a triangular form. A time-to-digital converter (TDC) records the arrival times and measures coincidences. Yellow lines correspond to single-mode fibers, the red dashed line to the free-space portion, and black lines to the electrical cables.}}
        \label{setup}
    \end{center}
\end{figure}

The experimental setup is shown in Fig.~\ref{setup}. The coherent state is emitted by an attenuated CW telecom laser. Its spectrum is represented by a Lorentzian function, centred at 1560.46 nm, with a width $\Delta \nu\approx~$10 kHz. To generate photon pairs, a CW laser is used to pump a periodically poled lithium niobate waveguide (PPLN/WG) at 780.23 nm. Note that these two lasers are independent. Pump photons are down-converted by spontaneous parametric down-conversion (SPDC) into photon pairs, say signal and idler, at the degenerate wavelength of 1560.46 nm. This non-linear process follows the laws of energy conservation and quasi-phase matching condition:
\begin{equation}
  \left\{
    \begin{array}{ll}
     \omega_{p} = \omega_{s} + \omega_{i} \\
     \textbf{k}_{p} = \textbf{k}_{s} + \textbf{k}_{i} - \frac{2\pi}{\Lambda} \textbf{n},
    \end{array}
    \right.  
\end{equation}
where $(\omega_{p,s,i},\textbf{k}_{p,s,i})$ the frequency and the wave vector of the pump, signal, and idler photons, respectively, $\Lambda$ the poling period of the PPLN, and $\textbf{n}$ a unit vector. The paired photons are filtered using a dense wavelength division multiplexer (DWDM) to select pairs close to the degenerate wavelength ($\omega_{s}\approx\omega_{i}$), and are then separated using a BS. One photon interferes with the coherent state, while the other acts as a timing trigger for the rest of the experiment, when detected by a SNSPD. The resulting detection signal triggers the CW telecom laser via an intensity modulator (IM), enabling a short detection time window ($\Delta t$=4~ns) and preventing saturation of the SNSPDs from the continuous photon flux, considering that photon pair generation is far less probable. We use a Time-to-digital converter (TDC) (see Fig.~\ref{setup}) to precisely adjust the trigger signal delivered to the IM, while also applying a bias tension to achieve a nearly perfect rectangular shape and narrow temporal width. A fiber spool is inserted along the signal path to compensate for travel delay and electro-optical conversion in the heralded arm path.

The photon pairs are created in a multimode fashion, as the SPDC process provides multimode temporal photons in the coherence time of the pump laser. To select only one temporal contribution, the coherence time of the photons must exceed the timing jitter of the overall detection system. To achieve this, a narrow spectral filter ($\Delta\nu=$540 MHz) is used to select only one temporal photonic contribution and has to satisfy the condition:
\begin{equation}\label{jitter}
     \frac{1}{\Delta\nu_{filter}}\gg \tau_{jitter},
     \label{Eq:equation}
\end{equation}
with $\Delta\nu_{filter}$ the spectral filter width and $\tau_{jitter}$ the convoluted timing jitter of the detectors, respectively. Here, $\frac{1}{\Delta\nu_{filter}}$ and $\tau_{jitter}$ are equal to 1900 and 150 ps, respectively. 

Another challenge lies in achieving spectral overlap between two incoming photons, each originating from a laser source and a spectrally filtered HSPS, with spectral bandwidths of a few tens of kHz and 540 MHz, respectively. This spectral overlap is managed through reciprocal filtering in the time domain, which is governed by the detection timing jitter. The effective spectrum of the two interfering photons is related to the inverse Fourier transform of the jitter, whose spectral bandwidth must exceed that of the photon filtering stage. This condition is the reciprocal of Eq.~\ref{Eq:equation} and is already met.

\begin{figure}[!ht]
    \begin{center}
        \includegraphics[width=0.9\linewidth]{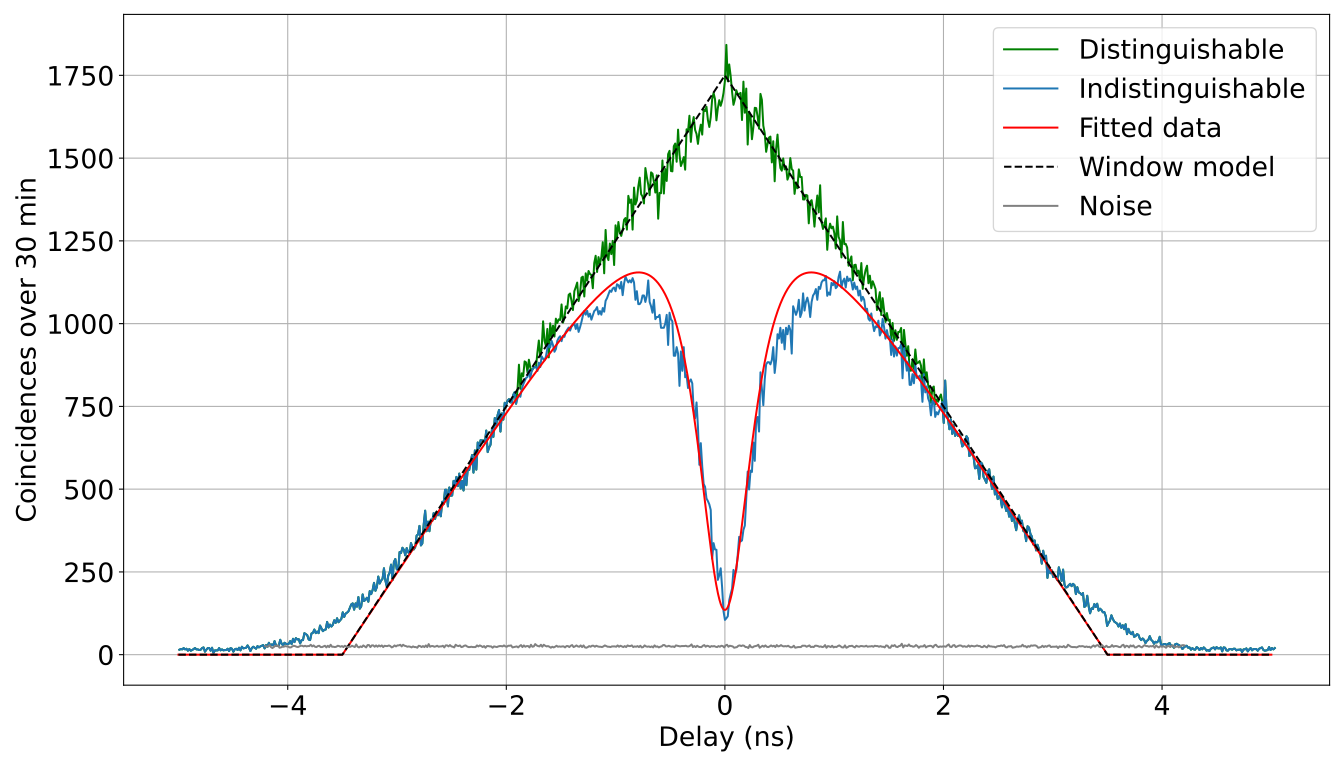}
        \caption{\small{Coincidence histograms recorded during 30 min as a function of temporal delay,$\tau$, between SNSPDs. \textit{Green curve}: data when photons are maximally distinguishable ($\Delta \nu = 6\,GHz$) ; \textit{blue curve}: data when photons are maximally indistinguishable $\Delta \nu = 0\,GHz)$; \textit{red curve}: fit based on experimental properties ; \textit{grey curve}: total noise ; \textit{dashed black line}: temporal window of the coincidence histogram.}}
        \label{dip}
    \end{center}
\end{figure}

Fig.~\ref{dip} shows the measured coincidence rate as a function of the delay $\tau$ between the two interfering photons. Heralded two-fold coincidences are registered by a TDC which stamps the detection events (t, t + $\tau$) and stores them in a table. We consider two cases: i) the photons at the BS are perfectly indistinguishable by adjusting all degrees of freedom, and ii) we introduce distinguishability between them. In our case, distinguishability is induced by slightly shifting the central wavelength of the WCS compared to that of the paired photons by more than one spectral mode, i.e., 540\,MHz. Specifically, we tune the telecom CW laser by 6\,GHz, thereby ensuring the distinguishability of the two spectral modes.
Furthermore, we have developed a theoretical model to fit the general shape of the HOM dip based on the temporal profiles of the detection windows and the spectral filter. As mentioned above, the first case, corresponding to "distinguishability" ($\Delta \nu \neq 0$), which arises from the convolution of two rectangular detection windows with a width of 4~ns. In this case, since the photons do not interfere, the related pattern has a shape of a triangle with a width of 8~ns (see "distinguishable fit" in Fig.~\ref{dip}). More interestingly, when the central wavelength of the spectral filter exactly matches that of the degenerate paired photons, the two interfering photons at the BS cause the center of the pattern to deepen. The shape and width of the HOM dip are influenced by the filter bandwidth. Our experimental data align well with our model, which predicts a full width at half maximum (FWHM) of 925 ps, while our measurement indicates 914 ps, resulting in a difference of less than 2\%. By fitting the data, without (with) the subtraction of accidental events, we obtain $V_{HOM,raw}$=90.7(5)\% and $V_{HOM,net}$ = 91.9(5)\%. The additional 1.2\% noise can mainly be attributed to the parasitic amplified spontaneous emission of the WCS and to the dark counts in the SNSPDs. In comparison to our model, the emission parameters of the sources have safely been kept low ($\alpha = \bar{n} \simeq 0.01$) to obtain a visibility above 90\%. For this set of parameters, the simulation returns $V_{HOM}(0.01,0.01) = 92.4\%$, showing the excellent match between the model and the experience.\\

\section{Discussion}
Numerous works have studied the HOM effect in different configurations related to quantum network scenarios. The current technical challenge lies in maintaining the indistinguishability of the photons prior to the BSM. We now aim at comparing our work with similar studies that report HOM visibilities, with the objective of evaluating the performance of our experiment relatively to the state-of-the-art. Among these studies, we may cite Refs.~\cite{valivarthi_teleportation_2020, takesue_quantum_2015, bussieres_quantum_2014}, which demonstrate visibilities of approximately 80\% between one photon (either from a pair or a coherent source) and a member of a photon pair. As a comparison, let us highlight three advantages of our approach : i) higher HOM visibility, which increases the prospect of greater fidelity for teleportation; ii) synchronization of photons by detection~\cite{halder2007entangling}, enabling an asynchronous regime; and iii) compatibility with the absorption bandwidth of current solid-state quantum memories (a few hundred MHz), thereby facilitating light-to-matter quantum teleportation based on telecom quantum memory or quantum dot spin qubits~\cite{saglamyurek_quantum_2015, thomas_deterministic_2023}. Finally, a recent article has reported near-perfect visibilities between one photon originating from a WCS and one member of a pair, while considering only a single master laser and thus not addressing the synchronization challenge~\cite{xu_characterization_2023}.\\
Going further, quantum teleportation with time-bin photonic qubits can be envisioned due to the high visibility reported here. To this end, identical unbalanced interferometers have to be added for generating superposition states and for enabling the projective measurement on Bell states at the BSM~\cite{samara2021entanglement}. With such a setup and our approach, the theoretical maximum teleportation fidelity would stand above 95\%. 

In summary, we have reported a two-photon experiment using two independent and distinct sources, resulting in a high-visibility HOM interference above 90\%. We have gone beyond the classical limit by ensuring high indistinguishability between photons, with temporal synchronization ensured by detection. Our model allows for HOM interference between a WCS and an HSPS, enabling us to estimate the optimal emission regime of the sources for high HOM visibility. Our experiment is fully fibered, therefore telecom-compatible, providing the opportunity to develop a real-field operational teleportation link with low noise and reasonable count rates. This achievement opens the way to exploring further possibilities and realizing time-bin teleportation while working with on-demand single-photon sources~\cite{somaschi_near-optimal_2016} to unlock new opportunities for even better performance.

\section*{Acknowledgment}
This work has been carried out within the framework of the OPTIMAL project, funded by the European Union through the European Regional Development Fund (ERDF). The authors also acknowledge financial support from the Agence Nationale de la Recherche (ANR) and the Agence d'Innovation de Défense (AID) under the Direction Générale de l'Armement (DGA) through the LIGHT project, as well as under the National Quantum Strategy via the PEPR-quantum project QCommtestbeds (ANR-22-PETQ-0011) and the Quantera project INQURE (ANR-22-QUA1-0002). Additional funding was provided by the French government through its Investments for the Future program under the Université Côte d'Azur UCA-JEDI project (Quantum@UCA), managed by the ANR (ANR-15-IDEX-01). 

\section*{Author information}
All the authors contributed equally to the entire process, from the first draft to the final version of the manuscript before submission. We all read, discussed, and contributed to the writing, reviewing, and editing. L.L., A.M. \& S.T.  coordinated and managed the project, ensuring its successful completion.

\section*{Competing interests}
The authors declare that there are no competing interests.

\section*{Data Availability}
Data are available from the authors on reasonable request.


\section*{References}

\begin{thebibliography}{10}

\bibitem{labonte2024integrated}
Laurent Labont{\'e}, Olivier Alibart, Virginia D’Auria, Florent Doutre, Jean
  Etesse, Gregory Sauder, Anthony Martin, {\'E}ric Picholle, and S{\'e}bastien
  Tanzilli.
\newblock Integrated photonics for quantum communications and metrology.
\newblock {\em PRX Quantum}, 5(1):010101, 2024.

\bibitem{munro_designing_2022}
William~J Munro, Nicolo'Lo Piparo, Josephine Dias, Michael Hanks, and Kae
  Nemoto.
\newblock Designing tomorrow's quantum internet.
\newblock {\em AVS Quantum Science}, 4(2), 2022.

\bibitem{PhysRevLett.98.230501}
Antonio Ac{\'\i}n, Nicolas Brunner, Nicolas Gisin, Serge Massar, Stefano
  Pironio, and Valerio Scarani.
\newblock Device-independent security of quantum cryptography against
  collective attacks.
\newblock {\em Physical Review Letters}, 98(23):230501, 2007.

\bibitem{Pironio_2009}
Stefano Pironio, Antonio Ac{\'\i}n, Nicolas Brunner, Nicolas Gisin, Serge
  Massar, and Valerio Scarani.
\newblock Device-independent quantum key distribution secure against collective
  attacks.
\newblock {\em New Journal of Physics}, 11(4):045021, 2009.

\bibitem{chen_quantum_2022}
Jiu-Peng Chen, Chi Zhang, Yang Liu, Cong Jiang, Dong-Feng Zhao, Wei-Jun Zhang,
  Fa-Xi Chen, Hao Li, Li-Xing You, Zhen Wang, et~al.
\newblock Quantum key distribution over 658 km fiber with distributed vibration
  sensing.
\newblock {\em Physical Review Letters}, 128(18):180502, 2022.

\bibitem{PhysRevLett.126.130502}
Stefan Langenfeld, Stephan Welte, Lukas Hartung, Severin Daiss, Philip Thomas,
  Olivier Morin, Emanuele Distante, and Gerhard Rempe.
\newblock Quantum teleportation between remote qubit memories with only a
  single photon as a resource.
\newblock {\em Physical Review Letters}, 126(13):130502, 2021.

\bibitem{pfaff_unconditional_2014}
Wolfgang Pfaff, Bas~J Hensen, Hannes Bernien, Suzanne~B van Dam, Machiel~S
  Blok, Tim~H Taminiau, Marijn~J Tiggelman, Raymond~N Schouten, Matthew
  Markham, Daniel~J Twitchen, et~al.
\newblock Unconditional quantum teleportation between distant solid-state
  quantum bits.
\newblock {\em Science}, 345(6196):532--535, 2014.

\bibitem{valivarthi_teleportation_2020}
Raju Valivarthi, Samantha~I Davis, Cristi{\'a}n Pe{\~n}a, Si~Xie, Nikolai Lauk,
  Lautaro Narv{\'a}ez, Jason~P Allmaras, Andrew~D Beyer, Yewon Gim, Meraj
  Hussein, et~al.
\newblock Teleportation systems toward a quantum internet.
\newblock {\em PRX Quantum}, 1(2):020317, 2020.

\bibitem{takesue_quantum_2015}
Hiroki Takesue, Shellee~D Dyer, Martin~J Stevens, Varun Verma, Richard~P Mirin,
  and Sae~Woo Nam.
\newblock Quantum teleportation over 100 km of fiber using highly efficient
  superconducting nanowire single-photon detectors.
\newblock {\em Optica}, 2(10):832--835, 2015.

\bibitem{xia_long_2017}
Xiu-Xiu Xia, Qi-Chao Sun, Qiang Zhang, and Jian-Wei Pan.
\newblock Long distance quantum teleportation.
\newblock {\em Quantum Science and Technology}, 3(1):014012, 2017.

\bibitem{samara_entanglement_2020}
Farid Samara, Nicolas Maring, Anthony Martin, Arslan~S Raja, Tobias~J
  Kippenberg, Hugo Zbinden, and Rob Thew.
\newblock Entanglement swapping between independent and asynchronous integrated
  photon-pair sources.
\newblock {\em Quantum Science and Technology}, 6(4):045024, 2021.

\bibitem{tanzilli_photonic_2005}
Sebastien Tanzilli, Wolfgang Tittel, Matthaeus Halder, Olivier Alibart, Pascal
  Baldi, Nicolas Gisin, and Hugo Zbinden.
\newblock A photonic quantum information interface.
\newblock {\em Nature}, 437(7055):116--120, 2005.

\bibitem{mcmillan_two-photon_2013}
AR~McMillan, L~Labont{\'e}, AS~Clark, B~Bell, O~Alibart, A~Martin,
  WJ~Wadsworth, S~Tanzilli, and JG~Rarity.
\newblock Two-photon interference between disparate sources for quantum
  networking.
\newblock {\em Scientific reports}, 3(1):2032, 2013.

\bibitem{rarity_high-visibility_1993}
JG~Rarity, J~Burnett, PR~Tapster, and R~Paschotta.
\newblock High-visibility two-photon interference in a single-mode-fibre
  interferometer.
\newblock {\em Europhysics Letters}, 22(2):95, 1993.

\bibitem{martin_quantum_2012}
Anthony Martin, Olivier Alibart, MP~De~Micheli, DB~Ostrowsky, and S{\'e}bastien
  Tanzilli.
\newblock A quantum relay chip based on telecommunication integrated optics
  technology.
\newblock {\em New Journal of Physics}, 14(2):025002, 2012.

\bibitem{PhysRevLett.104.137401}
Edward~B Flagg, Andreas Muller, Sergey~V Polyakov, Alex Ling, Alan Migdall, and
  Glenn~S Solomon.
\newblock Interference of single photons from two separate semiconductor
  quantum dots.
\newblock {\em Physical Review Letters}, 104(13):137401, 2010.

\bibitem{patel_two-photon_2010}
Raj~B Patel, Anthony~J Bennett, Ian Farrer, Christine~A Nicoll, David~A
  Ritchie, and Andrew~J Shields.
\newblock Two-photon interference of the emission from electrically tunable
  remote quantum dots.
\newblock {\em Nature photonics}, 4(9):632--635, 2010.

\bibitem{santori_indistinguishable_2002}
Charles Santori, David Fattal, Jelena Vu{\v{c}}kovi{\'c}, Glenn~S Solomon, and
  Yoshihisa Yamamoto.
\newblock Indistinguishable photons from a single-photon device.
\newblock {\em nature}, 419(6907):594--597, 2002.

\bibitem{PhysRevLett.85.290}
Christian Kurtsiefer, Sonja Mayer, Patrick Zarda, and Harald Weinfurter.
\newblock Stable solid-state source of single photons.
\newblock {\em Physical review letters}, 85(2):290, 2000.

\bibitem{beugnon_quantum_2006}
Jones Beugnon, Matthew~PA Jones, Jos Dingjan, Beno{\^\i}t Darqui{\'e},
  Ga{\"e}tan Messin, Antoine Browaeys, and Philippe Grangier.
\newblock Quantum interference between two single photons emitted by
  independently trapped atoms.
\newblock {\em Nature}, 440(7085):779--782, 2006.

\bibitem{felinto_conditional_2006}
Daniel Felinto, Chin-Wen Chou, Julien Laurat, EW~Schomburg, Hugues
  De~Riedmatten, and H~Jeff Kimble.
\newblock Conditional control of the quantum states of remote atomic memories
  for quantum networking.
\newblock {\em Nature Physics}, 2(12):844--848, 2006.

\bibitem{maunz_quantum_2007}
Peter Maunz, DL~Moehring, Steven Olmschenk, Kelly~Cooper Younge,
  DN~Matsukevich, and Christopher Monroe.
\newblock Quantum interference of photon pairs from two remote trapped atomic
  ions.
\newblock {\em Nature Physics}, 3(8):538--541, 2007.

\bibitem{Kim:20}
Heonoh Kim, Danbi Kim, Jiho Park, and Han~Seb Moon.
\newblock Hong--ou--mandel interference of two independent continuous-wave
  coherent photons.
\newblock {\em Photonics Research}, 8(9):1491--1495, 2020.

\bibitem{huwer_quantum-dot_2017}
J~Huwer, M~Felle, RM~Stevenson, J~Skiba-Szymanska, MB~Ward, I~Farrer, RV~Penty,
  DA~Ritchie, and AJ~Shields.
\newblock Quantum-dot based telecom-wavelength quantum relay.
\newblock {\em arXiv preprint arXiv:1704.07765}, 2017.

\bibitem{shields_entangled-led-driven_2016}
Christiana Varnava, R~Mark Stevenson, Jonas Nilsson, Joanna Skiba-Szymanska,
  Branislav Dzur{\v{n}}{\'a}k, Marco Lucamarini, Richard~V Penty, Ian Farrer,
  David~A Ritchie, and Andrew~J Shields.
\newblock An entangled-led-driven quantum relay over 1 km.
\newblock {\em Npj Quantum Information}, 2(1):1--7, 2016.

\bibitem{rarity_non-classical_2005}
JG~Rarity, PR~Tapster, and R~Loudon.
\newblock Non-classical interference between independent sources.
\newblock {\em Journal of Optics B: Quantum and Semiclassical Optics},
  7(7):S171, 2005.

\bibitem{bussieres_quantum_2014}
F{\'e}lix Bussi{\`e}res, Christoph Clausen, Alexey Tiranov, Boris Korzh,
  Varun~B Verma, Sae~Woo Nam, Francesco Marsili, Alban Ferrier, Philippe
  Goldner, Harald Herrmann, et~al.
\newblock Quantum teleportation from a telecom-wavelength photon to a
  solid-state quantum memory.
\newblock {\em Nature Photonics}, 8(10):775--778, 2014.

\bibitem{halder2007entangling}
Matth{\"a}us Halder, Alexios Beveratos, Nicolas Gisin, Valerio Scarani,
  Christoph Simon, and Hugo Zbinden.
\newblock Entangling independent photons by time measurement.
\newblock {\em Nature physics}, 3(10):692--695, 2007.

\bibitem{saglamyurek_quantum_2015}
Erhan Saglamyurek, Jeongwan Jin, Varun~B Verma, Matthew~D Shaw, Francesco
  Marsili, Sae~Woo Nam, Daniel Oblak, and Wolfgang Tittel.
\newblock Quantum storage of entangled telecom-wavelength photons in an
  erbium-doped optical fibre.
\newblock {\em Nature Photonics}, 9(2):83--87, 2015.

\bibitem{thomas_deterministic_2023}
SE~Thomas, L~Wagner, R~Joos, R~Sittig, C~Nawrath, P~Burdekin, T~Huber-Loyola,
  S~Sagona-Stophel, S~H{\"o}fling, M~Jetter, et~al.
\newblock Deterministic storage and retrieval of telecom quantum dot photons
  interfaced with an atomic quantum memory.
\newblock {\em arXiv preprint arXiv:2303.04166}, 2023.

\bibitem{xu_characterization_2023}
Aojie Xu, Lifeng Duan, Lirong Wang, and Yun Zhang.
\newblock Characterization of two-photon interference between a weak coherent
  state and a heralded single photon state.
\newblock {\em Optics Express}, 31(4):5662--5669, 2023.

\bibitem{samara2021entanglement}
Farid Samara, Nicolas Maring, Anthony Martin, Arslan~S Raja, Tobias~J
  Kippenberg, Hugo Zbinden, and Rob Thew.
\newblock Entanglement swapping between independent and asynchronous integrated
  photon-pair sources.
\newblock {\em Quantum Science and Technology}, 6(4):045024, 2021.

\bibitem{somaschi_near-optimal_2016}
Niccolo Somaschi, Valerian Giesz, Lorenzo De~Santis, JC~Loredo, Marcelo~P
  Almeida, Gaston Hornecker, S~Luca Portalupi, Thomas Grange, Carlos Anton,
  Justin Demory, et~al.
\newblock Near-optimal single-photon sources in the solid state.
\newblock {\em Nature Photonics}, 10(5):340--345, 2016.

\end{thebibliography}
\end{document}